\documentclass[aps,pre,twocolumn,superscriptaddress,showpacs]{revtex4}
\pdfoutput=1
\usepackage{graphicx,color,amssymb}

\begin{document}

\title{Phase transitions and marginal
ensemble equivalence for freely evolving flows\\
 on a rotating sphere}

\author{C. Herbert}
\email{corentin.herbert@lsce.ipsl.fr}
\affiliation{Service de Physique de l'Etat Condens\'e, DSM, CEA Saclay, CNRS URA 2464, 91191 Gif-sur-Yvette, France}
\affiliation{Laboratoire des Sciences du Climat et de l'Environnement, IPSL, CEA-CNRS-UVSQ, UMR 8212, 91191 Gif-sur-Yvette, France}
\author{B. Dubrulle}
\affiliation{Service de Physique de l'Etat Condens\'e, DSM, CEA Saclay, CNRS URA 2464, 91191 Gif-sur-Yvette, France}

\author{P.H. Chavanis}
\affiliation{Laboratoire de Physique Th\'eorique (IRSAMC), CNRS and UPS, Universit\'e de Toulouse, 31062 Toulouse, France}

\author{D. Paillard}
\affiliation{Laboratoire des Sciences du Climat et de l'Environnement, IPSL, CEA-CNRS-UVSQ, UMR 8212, 91191 Gif-sur-Yvette, France}

\begin{abstract}
The large-scale circulation of planetary atmospheres like that of the Earth is traditionally thought of in a dynamical framework. Here, we apply the statistical mechanics theory of turbulent flows to a simplified model of the global atmosphere, the \emph{quasi-geostrophic} model, leading to non-trivial equilibria. Depending on a few global parameters, the structure of the flow may be either a solid-body rotation (zonal flow) or a dipole. A second order phase transition occurs between these two phases, with associated spontaneous symmetry-breaking in the dipole phase. This model allows us to go beyond the general theory of marginal ensemble equivalence through the notion of Goldstone modes.
\end{abstract}


\maketitle

%% Intro
\section{Introduction}

It is well known that two-dimensional (2D) turbulent flows develop large-scale coherent structures \cite{McWilliams1984},
and widely believed that the existence of these structures is related to the existence of additional conserved quantities, as compared to the 3D case. In particular, Kraichnan \cite{Kraichnan1967} showed that in 2D flows, due to the conservation of enstrophy and energy, two inertial ranges coexist: a direct (downscale) enstrophy cascade and an inverse (upscale) energy cascade. As a result, the energy accumulates at the largest scales (lowest wavenumbers) leading to coherent structures. This phenomenon is sometimes related to a form of Bose-Einstein condensation \cite{Kraichnan1967,Tabeling2002}. The formation of these coherent structures can be accounted for by statistical mechanics approaches. Given some conserved quantities, one is able to predict the most probable flow, which, assuming ergodicity, coincides with the final state of the flow.
One particular area where large-scale coherent structures are ubiquitous, and dynamical approaches are both analytically intractable and numerically not affordable, is geophysical fluid dynamics. The early attempts to gain qualitative understanding of geophysical flows from statistical mechanics focused on the spectral space \cite{Salmon1976,Frederiksen1980}, but a theory dealing with continuous fields in the physical space is now available \cite{Miller1990,Robert1991a,Robert1991b}. The Miller-Robert-Sommeria (MRS) theory was used to develop small-scale parameterizations \cite{Robert1992,Kazantsev1998,Polyakov2001,Chavanis1997,Chavanis2002a,Chavanis2009} 
and study the formation of localized vortices like Jupiter's great red spot \cite{Michel1994a,Turkington2001,Bouchet2002}
or meso-scale ocean structures like vortex rings and zonal jets \cite{Weichman2006,VenailleArxiv}. It has also been discussed in relation to Fofonoff flows \cite{Venaille2009,Venaille2011,Naso2011}.

In this paper,
we construct the phase diagram of a simple model of atmospheric flow on a rotating sphere, the barotropic vorticity equation. Non-trivial equilibria are obtained and a second-order phase transition with spontaneous symmetry breaking is observed between a solid-body rotation phase and a dipole phase. In addition, the system displays surprising thermodynamic properties. Standard thermodynamics textbooks usually explain that for large enough systems (thermodynamic limit), mean values computed with a probability distribution that assigns equal weights to all the states having a given energy (microcanonical distribution) coincide with mean values computed with a Boltzmann distribution (canonical distribution). When the energy is not additive, which occurs in the presence of long-range interactions \cite{DauxoisLRIbook,Campa2009}, this is generally not the case. Consequently, peculiar thermodynamic features may appear: for instance negative microcanonical specific heat \cite{LyndenBell1968,Thirring1970} (due to the fluctuation-dissipation theorem, the canonical specific heat is always positive), convex dips in the microcanonical entropy \cite{Ellis2000}
and unusual phase transitions \cite{Bouchet2005,Chavanis2006h}.
The statistical equilibria of the barotropic vorticity equations on the rotating sphere have been the object of a number of studies in the past \cite{Ding2007,Lim2007a,MajdaWangBook}. However, no complete theory in the general framework of modern statistical mechanics of turbulent flows is available. The purpose of this paper is to fill this gap by providing the phase diagrams in the various statistical ensembles and discussing in detail a notion of marginal ensemble equivalence.

\section{Statistical Mechanics of Ideal Flows on a Rotating Sphere}

%% Description du systeme

We base our study on the quasi-geostrophic (QG) equations which provide a simple model of the atmospheric circulation. For simplicity, we shall treat here only the barotropic case with no bottom topography so that the potential vorticity is just the sum of the relative vorticity and the planetary vorticity: $q=\omega+f$, where $f=2\Omega \cos \theta$ and $\Omega$ is the angular velocity of the Earth. The general case will be described in detail elsewhere \cite{Herbert2011d}.
In the absence of forcing and dissipation, the potential vorticity is advected by the flow:
\begin{equation}\label{qgeqs}
\partial_t q + {\bf u} \cdot \nabla q =0.
\end{equation}
Here ${\bf u}=-\hat{\bf r}\times \nabla\psi$ is the velocity field and $\psi$ the stream function, related to the vorticity field by $\omega=-\Delta \psi$. For convenience, we adopt the Gauge condition $\int_{\cal D}\psi\, d{\bf r}=0$. Similarly to the 2D Euler equations, QG flows admit dynamical invariants: the energy $E[q]=(1/2) \int_{\cal D} (q-f)\psi\, d{\bf r}$ and the Casimirs $\Gamma_n[q]=\int_{\cal D} q^n\, d{\bf r}$. For specific flow domains ${\cal D}$, some additional invariants must be considered. Here, we focus on the case of a sphere with unit radius ${\cal D}=S^2$, for which the angular momentum $L[q]=\int_{\cal D} (q-f)\cos \theta\, d{\bf r}$ is conserved.

Steady-states of the QG equations which are stationary in a frame rotating with angular velocity $\Omega_L$ with respect to the terrestrial frame satisfy $q(\theta,\phi,t)=q(\theta,\phi-\Omega_L t)$. Substituting this relation into the evolution equation (\ref{qgeqs}), we obtain $\{q,\psi+\Omega_L \cos \theta\}=0$ where $\{,\}$ denote the Poisson brackets. Hence the general form of steady states of the QG equations on the sphere is $q=F(\psi+\Omega_L \cos \theta)$ where $F$ is an arbitrary function. Statistical mechanics allows us to select the most probable function $F$ given the value of the dynamical invariants $E,L$ and $\Gamma_n$ as follows.

%% Mecanique statistique

A priori, the evolution of the flow is purely determined by the initial potential vorticity field $q_0$. It is well known that throughout the dynamical evolution, the potential vorticity field develops finer and finer filaments. This small-scale behaviour of the potential vorticity field makes it difficult to treat the flow in a deterministic way, as the typical scale of potential vorticity fluctuations will eventually become smaller than any simulation or observation resolution. On the other hand, the large-scale structure can be readily predicted by statistical mechanics. Introducing a coarse-grained potential vorticity field, one maximizes the statistical entropy subject to the conservation constraints to obtain the most probable potential vorticity field \cite{Miller1990,Robert1991a,Robert1991b}. It is characterized by a functional relationship $q=F(\psi+\Omega_L \cos \theta)$ where the function $F$ depends on the invariants. In the following, we shall assume that $F$ is linear.  Several justifications can be given to motivate this choice: (i) it corresponds to a MRS equilibrium with a Gaussian potential vorticity distribution reached for particular initial conditions \cite{Miller1990}. Furthermore, a state that maximizes $S[q]=-(1/2)\int_{\cal D} q^2\, d{\bf r}$ at fixed energy, circulation and angular momentum is granted to be thermodynamically stable \cite{Bouchet2008,Venaille2009} (see also \cite{Chavanis2009}); (ii) a linear $q-\psi$ relationship is obtained in the limit of strong mixing and in the low energy limit \cite{Chavanis1996}; (iii) maximizing the statistical entropy with conservation of energy, circulation, angular momentum and fine-grained enstrophy is equivalent to minimizing the coarse-grained enstrophy $(1/2)\int_{\cal D} q^2\, d{\bf r}$ defined in terms of the coarse-grained potential vorticity at fixed energy, circulation and angular momentum \cite{Naso2010a}. The ensemble-mean potential vorticity is characterized by  a linear $q-\psi$ relationship and the fluctuations around it are Gaussian; (iv) some authors have proposed to treat the Casimirs in a canonical way \cite{Ellis2002}, which is equivalent to choosing a prior small-scale potential vorticity distribution \cite{Ellis2002,Chavanis2008b,Chavanis2010c}. The ensemble-mean potential vorticity then maximizes a generalized entropy $S[q]=-\int_{\cal D} C(q)\, d{\bf r}$ where $C$ is a convex function related to the prior. A  Gaussian prior leads to a linear $q-\psi$ relationship; it is associated with a quadratic generalized entropy $S[q]=-(1/2)\int_{\cal D} q^2\, d{\bf r}$ proportional to the coarse-grained potential enstrophy. All these justifications lead to a linear $q-\psi$ relationship. However, they are not equivalent. They essentially differ in the manner that the fragile constraints (high-order Casimir invariants) are treated, while the robust constraints (energy, circulation, angular momentum) are strictly taken into account in all these approaches.  In approaches (i) and (ii), although the structure of the equilibria in the limit considered depends only on the robust invariants, conservation of the fragile invariants is taken into account implicitly. On the contrary, approaches (iii) and (iv) are based on the idea that, in real flows, forcing and dissipation acting at small scales break down the conservation of high-order Casimirs. In (iii) these constraints are purely  discarded while in (iv) they are replaced by a prior.  Note that strictly speaking, either we consider the unforced-undamped case and we should consider all the invariants, or we consider the forced-damped case in which case the value of the robust invariants should also be selected by the forcing and the dissipation. Finally, the approach (ii) does not make any assumption on the fluctuations of the potential vorticity, contrary to (i), (iii) and (iv) in which they are Gaussian.

The microcanonical variational problem reads
\begin{equation}\label{mcvarprobeq}
{\cal S}(E,L)=\max_q \{ S[q] | E[q]=E, L[q]=L\}.
\end{equation}
Note that it is not necessary to include the circulation in the constraints as it always vanishes due to the geometry. The solutions of this variational problem are a subset of the full class of MRS equilibria.
In practice, constrained variational problems like (\ref{mcvarprobeq}) are difficult to solve. It is much more convenient to work with the dual variational problem with relaxed constraints, the grand-canonical variational problem:
\begin{equation}\label{gcvarprobeq}
{\cal J}(\beta,\mu)=\max_q \{ S[q]-\beta E[q]-\mu L[q]\}.
\end{equation}
Clearly the two variational problems have the same critical points, but the nature of the critical points (maxima, minima, saddle points) may differ. When this happens, we speak of \emph{ensemble inequivalence} \cite{Ellis2000,Ellis2002,Touchette2004}.
For isolated systems, like the large systems encountered in astrophysics or geophysical fluid dynamics, the natural ensemble is the microcanonical ensemble. The classical interpretation of the grand-canonical ensemble is that the system is in contact with a reservoir of heat and angular momentum. The physical interpretation of such a reservoir is unclear. Nevertheless, it is always worthwhile to consider the grand-canonical ensemble, be it just as a useful mathematical device.
Indeed, a solution of the grand-canonical problem (\ref{gcvarprobeq}) is always a solution of the more constrained dual microcanonical problem (\ref{mcvarprobeq}).  Note that the variational
principles (\ref{mcvarprobeq}) and (\ref{gcvarprobeq}) also provide
sufficient conditions of nonlinear dynamical stability for the QG
equations \cite{Ellis2002} (see also \cite{Chavanis2009}).

\section{Solutions of the mean-field equation and phase transition}

The critical points of (\ref{mcvarprobeq}) and (\ref{gcvarprobeq}) satisfy $\delta S - \beta \delta E - \mu \delta L=0$, which yields a linear $q-\psi$ relationship: $q=-\beta \psi - \mu \cos \theta$, in accordance with the general form for steady-states $q=F(\psi+\Omega_L \cos \theta)$, where $\Omega_L=\mu/\beta$ is the angular velocity of the frame in which the flow is stationary. Replacing $q$ by its definition, we obtain the mean-field equation 
\begin{equation}\label{dipoleeq}
\Delta \psi - \beta \psi=f+\mu \cos \theta.
\end{equation}
%% Etude de l'equation champ moyen
To solve this Helmholtz equation, we introduce the eigenvalues and eigenvectors of the Laplacian on the sphere: $\Delta Y_{nm} = \beta_n Y_{nm}$, where the spherical harmonics $Y_{nm}$ form an orthonormal basis of the Hilbert space $L^2({\cal D})$, and $\beta_n=-n(n+1)$ ($n \in \mathbb{N}$, $-n \leq m \leq n$).

When $\beta$ is not an eigenvalue of the Laplacian, the operator $\Delta-\beta I$ is invertible and the equilibrium stream function reads
\begin{equation}\label{dipoleeq}
\psi= \Omega_* \cos \theta,
\end{equation} 
with $\Omega_*=(2\Omega+\mu)/(\beta_1-\beta)$. This corresponds to a solid-body rotation with angular velocity $\Omega_*$. The energy, angular momentum and entropy depend only on the angular velocity: $E={\Omega_*}^2/3, L=2\Omega_*/3, S=-2(\Omega+\Omega_*)^2/3$. In particular, for a solid-body rotation, the energy and angular momentum are not independent conserved quantities: they are related by $E=E_*(L)$ where $E_*(L)=3L^2/4$. One can easily show that the energy of any flow with angular momentum $L$ is always greater than the energy of the solid-body rotation with the same angular momentum: $E \geq E_*(L)$.

When $\beta$ is an eigenvalue of the Laplacian but not the first non-zero one - say $\beta=\beta_{n>1}$, the stream function belongs to a $2n+1$ dimensional vector space: $\psi = \Omega_* \cos \theta + \sum_m \psi_{nm} Y_{nm}$ with again $\Omega_*=(2\Omega+\mu)/(\beta_1-\beta_n)$. It is a superposition of a solid-body rotation with a multipole. The energy, angular momentum and entropy are found to be $E={\Omega_*}^2/3-\beta_n \sum_m |\psi_{nm}|^2/2, L=2\Omega_*/3, S= -2(\Omega+\Omega_*)^2/3-\beta_n \sum_m | \psi_{nm}|^2/2$.

\begin{figure}
\includegraphics[width=\linewidth]{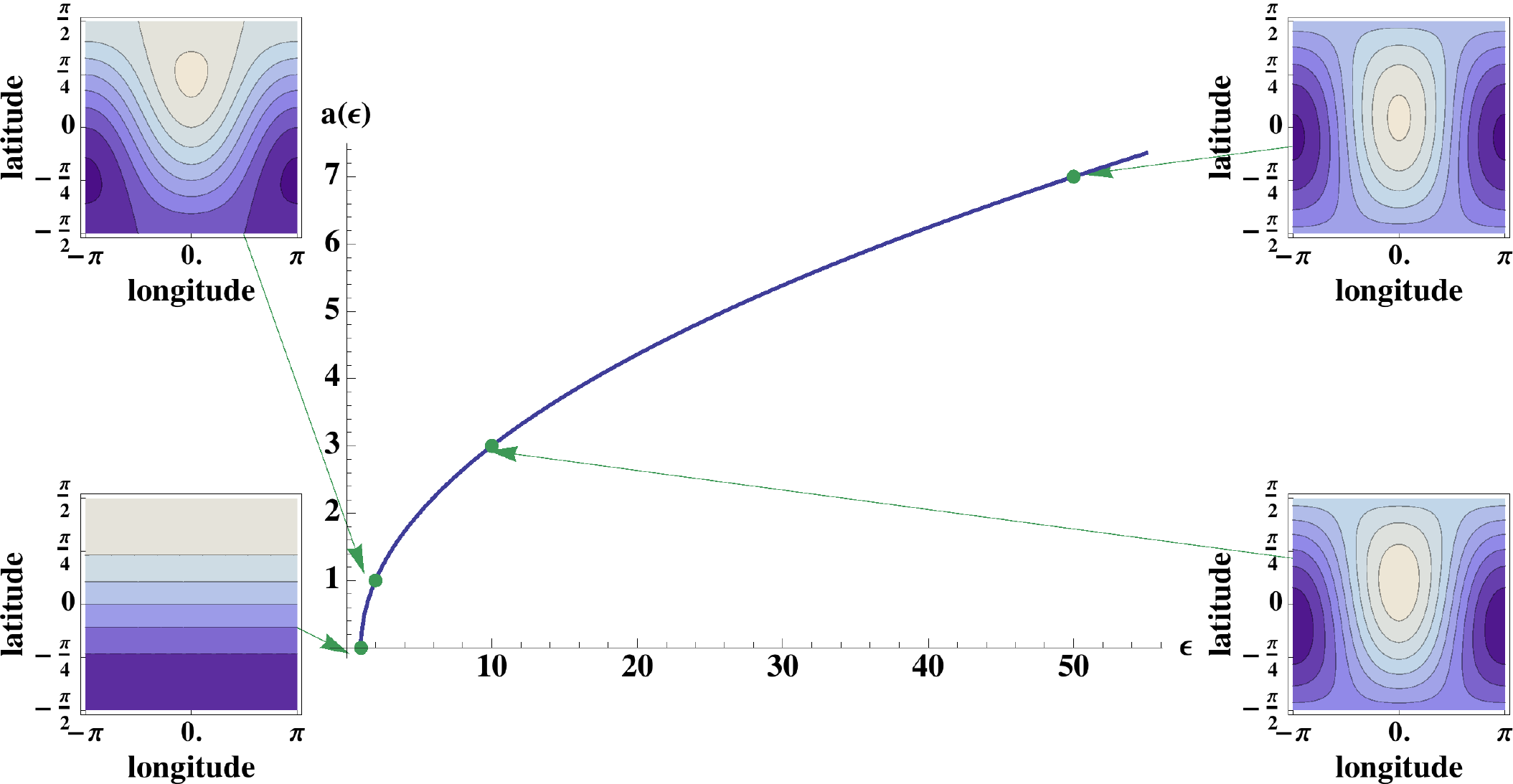}
\caption{(color online). Amplitude of the dipole, relative to the solid-body rotation, as a function of the control parameter $\epsilon$. At $\epsilon=\epsilon_c\equiv 1$, a second-order phase transition occurs, accompanied by spontaneous symmetry breaking (in the dipole phase). Sample stream functions are shown in the insets, with the arbitrary phase $\phi_0$ set to $0$. }\label{dipoleplot}
\end{figure}

Finally, the mean-field equation admits solutions for $\beta=\beta_1$ only if  $\mu=\mu_c\equiv-2\Omega$. In that case, the stream function reads $\psi=\Omega_* \cos \theta + \gamma_c \sin \theta \cos \phi + \gamma_s \sin \theta \sin \phi$. The energy, angular momentum and entropy depend on the coefficients $\Omega_*,\gamma_c,\gamma_s$: $E=({\Omega_*}^2+\gamma_c^2+\gamma_s^2)/3, L=2 \Omega_*/3, S= - 2((\Omega+\Omega_*)^2+\gamma_c^2+\gamma_s^2)/3$. Introducing the angle $\phi_0$ such that $\gamma_c=\sqrt{3(E-E_*(L))} \cos \phi_0$ and $\gamma_s=\sqrt{3(E-E_*(L))}\sin \phi_0$, the stream function becomes
\begin{equation}\label{dipoleeq}
\psi=\Omega_* \cos \theta + \sqrt{3(E-E_*(L))} \sin \theta \cos (\phi-\phi_0).
\end{equation}
This is a dipole flow with an arbitrary phase $\phi_0$. Setting $\epsilon=E/E_*(L)$, the amplitude of the dipole relative to the background solid-body rotation can be recast as $a(\epsilon)\equiv \sqrt{\epsilon-1}$; then $\psi = (3L/2)\lbrack \cos \theta + a(\epsilon) \sin \theta \cos (\phi-\phi_0) \rbrack$. This reveals a second-order phase transition between a solid-body rotation phase at $\epsilon=\epsilon_c \equiv 1$ and a dipole phase for $\epsilon> \epsilon_c$ (Fig. \ref{dipoleplot}). The angular momentum controls the overall amplitude. The phase $\phi_0$ is not determined by the control parameters: this is a case of spontaneous symmetry breaking.
Since the phase transition line coincides with the line of minimum energy $E_*(L)$, the phase transition would be difficult to see in practice (in numerical simulations for instance). This is reminiscent of some systems encountered in condensed matter physics, where the critical point is sometimes reached at vanishing temperature ($T=0$), like for instance in the 1D Ising chain. However, it is possible that,  in more realistic models, the two lines separate from each other.

We stress that the spontaneous symmetry-breaking property is also related to an apparent paradox of the Kraichnan statistical theory in spectrally truncated space \cite{Kraichnan1975}. In the spectrally truncated statistical mechanics of quasi-geostrophic flows over topography \cite{Salmon1976,Frederiksen1980,Carnevale1987}, there is usually a distinction between the mean flow and the eddies, or \emph{transients}. Statistical equilibria are said to be \emph{sharp} when the eddy component vanishes in the limit of infinite cut-off wavenumber. An often-read claim is that in the absence of topography, all the energy is contained in the transients and there is no mean flow in the statistical equilibria, breaking the sharpness property. This would be in contradiction with the MRS theory which directly predicts \emph{sharp} equilibria. This apparent paradox is waived by a symmetry breaking argument: for a given value of the canonical parameters, opposite stream functions are equally admissible. When summing over all the states, they cancel out, leaving a null mean flow. However in practice, the system selects spontaneously a particular symmetry-breaking state within the symmetry restoring set of equilibrium states. Note that this apparent paradox is more easily understood in the context of inequivalence of statistical ensembles: the proper ensemble to work with is the microcanonical ensemble, in which the paradox does not appear. But this was not noticed at that time because statistical mechanics in the microcanonical ensemble was still poorly understood (the first observation of ensemble inequivalence was made in \cite{LyndenBell1968,Thirring1970}, but it took some time before a thorough understanding was reached \cite{Ellis2000}), although Kraichnan himself expressed concerns about the relevance of canonical measures for fluid systems.

\section{Nature of the critical points and marginal ensemble equivalence}
%% Stability of the critical points

The nature of the critical points of entropy is given by the second variations of the grand-potential functional $J[q]\equiv S[q]-\beta E[q] -\mu L[q]$: a critical point of entropy at fixed energy and angular momentum is a local maximum if and only if $\delta^2 J <0$ for all perturbations $\delta q$ that conserve $E$ and $L$ at first order \cite{Chavanis2009}. This is the microcanonical stability condition. In the grand-canonical ensemble, the stability condition is $\delta^2 J <0$ for all perturbations. The quadratic form $\delta^2 J = -(1/2)\int_{\cal D} (\delta q)^2\, d{\bf r} - (1/2)\beta \int_{\cal D} (\nabla \delta \psi)^2\, d{\bf r}$ is negative definite for $\beta>\beta_1$. Hence, in this case, the flow is grand-canonically stable and therefore also microcanonically stable. When $\beta<\beta_1$, it is easy to build perturbations that destabilize the flow. Consider a perturbation $\delta q$ proportional to a Laplacian eigenvector $Y_{nm}$. The basic flow is either a solid-body rotation, or a degenerate flow ($\beta=\beta_{p>1}$). For $(n,m)\neq (0,0),(1,0)$ (and $n\neq p$ in the degenerate case), the perturbation conserves the energy and the angular momentum, while at the same time $\delta^2 J >0$. Thus, all the states obtained for $\beta<\beta_1$ are microcanonically unstable, and therefore also grand-canonically unstable \cite{Herbert2011d}.

When $\beta=\beta_1$ the situation is slightly more subtle: the quadratic form $\delta^2 J$ is degenerate: its radical $R$ is a vector space of dimension 3, spanned by $Y_{11},Y_{1,-1}$ and $Y_{10}$. All the elements of this vector space (which are combinations of solid-body rotations and dipoles) share the same value $J=-2\Omega^2/3$ of the grand-potential functional, which means that they are metastable in the grand-canonical ensemble. Perturbations belonging to $R$ allow transitions between one flow of the form (\ref{dipoleeq}) to another with different values of $L,E$ and $\phi_0$. In the microcanonical ensemble, imposing conservation of angular momentum and energy at all orders implies that the only perturbations in $R$ to be considered are those which only change $\phi_0$. These solutions have the same entropy ${\cal S}=-2(E-E_*(L))-(3/2)(L+2\Omega/3)^2$. Hence each state with given $(E,L)$ in the microcanonical ensemble is metastable, but with fewer \emph{Goldstone bosons} (only one) than in the grand-canonical ensemble (where there are three Goldstone bosons).

%% Proprietes thermodynamiques

\begin{figure}
\includegraphics[width=\linewidth]{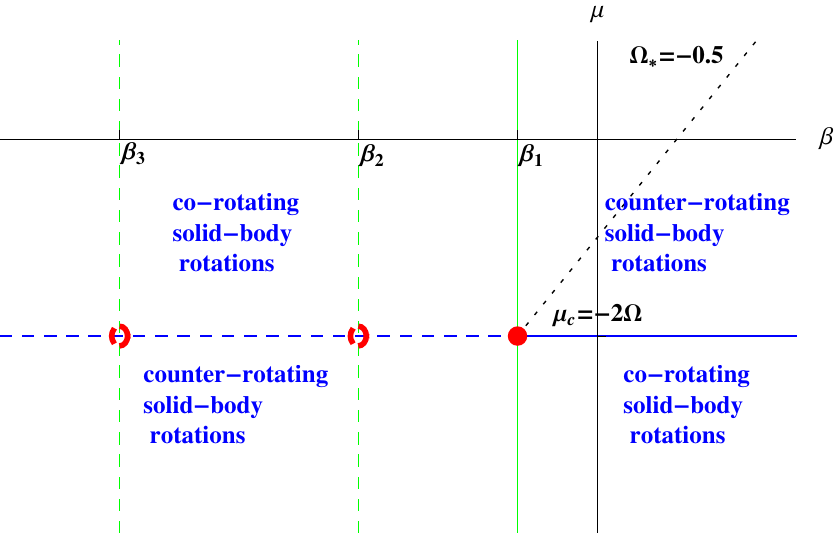}
\caption{(color online). Grand-canonical phase diagram for barotropic flow on a rotating sphere. Stable equilibrium states are obtained for $\beta>\beta_1$. At the critical point $(\beta=\beta_1,\mu=\mu_c)$, there are an infinity of metastable states. Vertical dashed lines (green) indicate the position of unstable degenerate states. At the intersection with the line $\mu=\mu_c$, (dashed red points), only the degenerate part subsists. On the horizontal blue solid (resp. dashed) line the equilibrium flow is a stable (resp. unstable) trivial motionless flow. No solution exists if $\beta=\beta_1$ and $\mu\neq \mu_c$. The dotted half straight-line represents a iso-$\Omega_*$ line.}\label{phasediagramGCfig}
\end{figure}

\begin{figure}
\includegraphics[width=\linewidth]{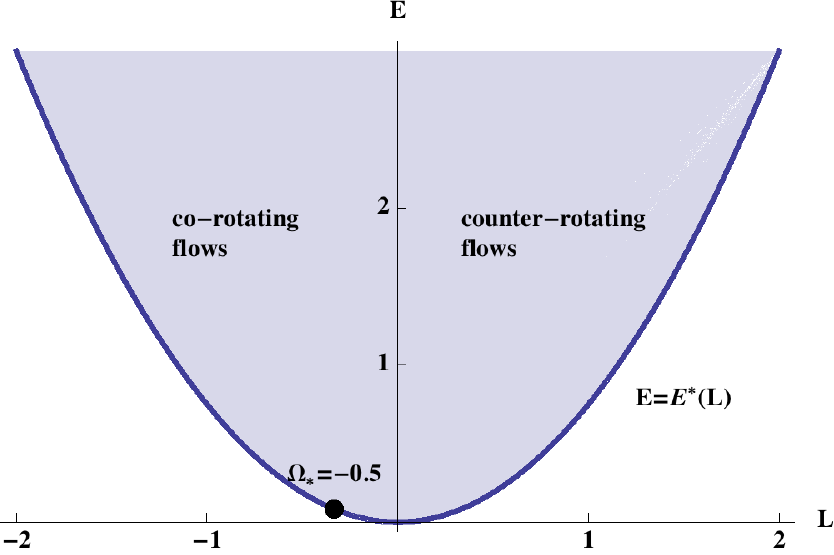}
\caption{(color online). Microcanonical phase diagram for barotropic flows on a rotating sphere. The parabola $E=E_*(L)$ corresponds to solid-body rotations, while the shaded area over the parabola represents dipole flows. The area below the parabola is not accessible. The parabola is the location of a second-order phase transition with spontaneous symmetry-breaking.}\label{phasediagramMCfig}
\end{figure}

To sum up, in the grand-canonical ensemble, the equilibrium flow is either a solid-body rotation ($\beta>\beta_1$, co-rotating for $\mu<\mu_c$, counter-rotating for $\mu>\mu_c$) or a dipole ($\beta=\beta_1, \mu=\mu_c$) with arbitrary energy, angular momentum and phase. The corresponding phase diagram is represented on Fig. \ref{phasediagramGCfig}. In the particular case $\mu=0$ (no angular momentum constraint), we recover the solid-body rotations obtained previously in Monte-Carlo simulations \cite{Ding2007}.
In the microcanonical ensemble, when $E=E_*(L)$ the equilibrium flow is a solid-body rotation (co-rotating for $L>0$, counter-rotating for $L<0$), while for $E>E_*(L)$ it is a dipole with arbitrary phase (Fig. \ref{phasediagramMCfig}). A second-order phase transition occurs between the solid-body-rotation phase and the dipole phase, accompanied by $U(1)$ spontaneous symmetry breaking, as the phase of the dipole is undetermined. The choice of a particular phase is analogous to the choice of a \emph{fundamental state} in field theory \cite{ZinnJustinBook}. 

The second order phase transition is very special because one of the two phases (solid-body rotation) exists only on a curve $E_*(L)$ while the other phase (dipole) exists on a surface $E>E_*(L)$. On the curve $E_*(L)$, the parameters $\beta(E,L)$ and $\mu(E,L)$ take any values in the range $\beta\ge \beta_1$  and $\mu<\mu_c$ (if $L>0$) or $\mu>\mu_c$ (if $L<0$), while on the surface $E>E_*(L)$ they take the unique values  $\beta=\beta_1$ and $\mu=\mu_c$. Therefore, the second derivatives $\partial^2{\cal S}/\partial E^2=\partial \beta/\partial E$ and  $\partial^2 {\cal S}/\partial L^2=\partial\mu/\partial L$ pass from $+\infty$ to $0$ as we go from the solid-body rotation phase to the dipole phase. This is therefore a case of extreme discontinuity of the second derivatives of the entropy with respect to the conserved quantities.

Note that the analytical tractability of the system considered here allows us to understand precisely how the grand-canonical equilibrium states and the microcanonical equilibrium states are related: to every half-straight line starting from the critical point on the grand-canonical phase diagram (in the right half-plane, see Fig. \ref{phasediagramGCfig}), with equation $\mu=\Omega_*(\beta_1-\beta)-2\Omega$, we can associate a single point $(E,L)=(\Omega_*^2/3,2\Omega_*/3)$ on the $E=E_*(L)$ parabola in the microcanonical phase diagram (Fig. \ref{phasediagramMCfig}). Similarly, the critical point $(\beta=\beta_1,\mu=\mu_c)$ is mapped onto the whole area over the parabola. Formally, the grand-canonical and microcanonical ensembles are equivalent at the macrostate level \cite{Touchette2004}; it is in fact a case of marginal equivalence \cite{Ellis2000}. Note that the marginal ensemble equivalence is extreme here in that the vast majority of the equilibrium states in the microcanonical ensemble ($E>E_*(L)$) are all included in the set of grand-canonical equilibria obtained at a single value of the Lagrange parameters $(\beta=\beta_1, \mu=\mu_c)$. As a consequence, the construction of equilibrium states in the grand-canonical ensemble may be difficult to control in practice.
Marginal ensemble equivalence is also seen at the thermodynamic level: geometrically, the microcanonical entropy ${\cal S}$ is a plane. Therefore, it is a concave function, which is characteristic of ensemble equivalence at the thermodynamic level \cite{Ellis2000,Touchette2004}, but only marginally so: it is also a convex function. Equivalently, the specific heats $\partial^2 {\cal S}/\partial E^2$ and $\partial^2 {\cal S}/\partial L^2$ vanish.
This indicates that the system considered is on the edge of ensemble inequivalence: the grand-canonical and microcanonical ensembles are formally equivalent, but very close to being inequivalent.

Here it is possible to analyze the ensemble equivalence properties beyond marginal ensemble equivalence. Indeed, as explained above, the metastability in the grand-canonical ensemble is more severe than in the microcanonical ensemble since only the phase of the dipole is undetermined (and thus free to vary due to spontaneously generated perturbations) in the microcanonical ensemble while the energy and angular momentum are also unconstrained in the grand-canonical ensemble. Hence a refinement of the notion of marginal ensemble equivalence can be introduced by considering the number of Goldstone modes in each ensemble. It is equal to one in the microcanonical ensemble, three in the grand-canonical ensemble, and two in any mixed ensemble (not discussed). As a practical manifestation of these Goldstone modes, numerical constructions of the equilibrium states have to be aware that spontaneous fluctuations may exist, the form of which depends on the statistical ensemble considered. This property stems from the degeneracy of the first eigenvalue of the Laplacian on the sphere. Up to now, most studies have assumed that the first Laplacian eigenmode was not degenerate \cite{Venaille2009,Venaille2011,Bouchet2010,CorvellecThesis}, with the notable exception of \cite{MajdaWangBook,Bouchet2012}. Another important consequence is that in the presence of degeneracies, ensemble inequivalence results such as those observed in \cite{Venaille2009} can collapse.

Note also that the marginal ensemble equivalence and the second order phase transition observed here may not be robust with respect to small nonlinearities in the $q - \psi$ relationship. As analyzed in the case of the energy-circulation ensemble on a rectangular domain \cite{CorvellecThesis}, depending on the sign of the small nonlinearity, second order phase transitions can turn into first order phase transitions. Due to the formal similarity between the two systems, the situation is likely to be the same here.

% Remarques finales
\section{Conclusion}

In this paper, we have explicitly computed  a class of statistical equilibria for 2D flows on a rotating sphere, built the corresponding phase diagrams and characterized precisely the ensemble equivalence properties. Remark that all the energy condensates
in the lowest (non-zero) order mode. It is the combination of the angular momentum conservation and Laplacian degeneracy on the sphere that allows for non-trivial Bose-Einstein condensation \cite{Kraichnan1967}; with the angular momentum constraint, the energy is shared between two phases, which makes possible phase coexistence and phase transition. Hence the results presented here have a fundamentally geometric nature. Note that in the dipole phase, the value of the angular momentum fixes the solid-body rotation part of the flow.
The subtle ensemble equivalence properties of the system also stem from the particular geometry: the general inequality $E\geq E_*(L)$ is purely kinematic. Without this inequality, ensemble equivalence would not hold. Nevertheless, ensemble equivalence is only marginal \cite{Ellis2000}, and it is so in a particularly extreme way. At the thermodynamic level, the entropy surface is concave, but not strictly: it is locally flat at each point of the interior of the parabola $E=E_*(L)$. At the macrostate level, each microcanonical equilibrium state is also a grand-canonical equilibrium state. More precisely, the set of macrostates ${\cal M}(E,L)$ obtained in the microcanonical ensemble with a given energy $E$ and angular momentum $L$  is a subset of the set of macrostates ${\cal GC}(\beta,\mu)$ obtained in the grand-canonical ensemble with the adequate Lagrange parameters $(\beta,\mu)$. When $E=E_*(L)$,  ${\cal M}(E,L) = {\cal GC}(\beta,\mu)$ for any $(\beta,\mu)$ satisfying $(2\Omega+\mu)/(\beta_1-\beta)=\Omega_*$ with $\Omega_*=3L/2$. For $E>E_*(L)$, ${\cal M}(E,L) \subsetneq {\cal GC}(\beta_1,\mu_c)$. Indeed, the set ${\cal GC}(\beta_1,\mu_c)$ contains all the dipole flows with any value of the energy and angular momentum in the interior of the parabola, while the set ${\cal M}(E,L)$ only contains the dipole flows with given values of the energy and angular momentum.
This is a loose constraint on the two macrostate sets. We have shown that it is possible to go further than the usual notion of marginal ensemble equivalence  by introducing the concept of symmetry breaking: the symmetry breaking occurring in the grand-canonical ensemble is stronger than in the microcanonical ensemble. As usual in the theory of continuous symmetry breaking, this is measured by the number of Goldstone modes. Statistical mechanics in the microcanonical and grand canonical ensembles differ in their number of such modes (zero-energy modes in the microcanonical ensemble and zero-temperature modes in the grand canonical ensemble). This subtle difference can exist only when the Laplacian is degenerate on the flow domain, which is the case on the sphere and on any manifold with sufficient symmetries (in fact, it suffices that the eigenspace corresponding to the first non-zero eigenvalue of the Laplacian be degenerate). For instance, the analysis does not apply to the case of a rectangular domain with linear $q-\psi$ relationship and zero circulation \cite{Chavanis1996, Venaille2011}, even though the entropy depends linearly on the energy in that case. Contrary to our study, symmetries are not sufficient in this example to go further than the notion of marginal ensemble equivalence: on a rectangular domain, one only has a discrete, not continuous, symmetry breaking, which prevents the discussion in terms of Goldstone bosons. This explains why it was not discussed before, as all authors focused on cases when the first eigenvalue of the Laplacian is not degenerate. Therefore, there exist cases for which a refinement of the definition of marginal ensemble equivalence is needed. As Laplacian degeneracies stem from symmetries, it is natural that the degree of symmetry breaking in macrostates becomes the refined notion of marginal ensemble equivalence at the macrostate level.
In our system, we are able to say that there is one Goldstone mode in the microcanonical ensemble, three Goldstone modes in the grand-canonical ensemble and two Goldstone modes in both mixed ensembles (where one constraint is treated microcanonically and the other canonically).

Note that in the case of a rectangular domain with doubly-periodic boundary condition (the two-dimensional torus), the first Laplacian eigenmode is degenerate (for aspect ratio one). The consequence is that the topology of the flow is not fixed by the invariants when considering only the maximization of a quadratic generalized entropy functional with energy constraint: both parallel flows and dipole flows belong to the family of extrema \cite{Bouchet2012}. The degeneracy can be lifted either by stretching the domain or by considering small non-linearities in the $q-\psi$ relationship. The coexistence of two phases is very similar to our findings in the spherical geometry, except that in our case, even at fixed energy, the degeneracy can be lifted by the angular momentum constraint (in the microcanonical ensemble). The square doubly periodic domain is another example of a case of marginal ensemble equivalence where our analysis in terms of Goldstone mode is expected to be valid, with the difference that the marginal ensemble equivalence there only holds with respect to one parameter, the energy, whereas in the spherical case it holds in a two-dimensional thermodynamic parameter space, the $(E,L)$ plane.

From the geophysical application side, our study is based on the QG equations. These equations provide a simplified model of the true equations of motion for geophysical fluids, which are often called ``primitive equations'' and rely on the Navier-Stokes equations. In the QG approximation, the essential hypotheses are that the fluid layer is thin and that geostrophic balance prevails approximately (which implies that rotation is strong enough) \cite{PedloskyGFD}. Both hypotheses are relatively well verified in the atmosphere and oceans. Hence, although we consider here the simplest version of the QG equations (barotropic case), the model in itself is relatively realistic (it is not a toy model). However, there is of course a fundamental difference between the model and reality, as we consider here that the forcing and dissipation equilibrate locally while it is only true globally in real steady states. This strong simplification is common to all the statistical mechanics approaches up to now, as it ensures that the Liouville theorem is satisfied.

It is generally believed that the dynamics of the atmosphere is essentially due to out-of-equilibrium processes. Nevertheless, the equilibrium states obtained here resemble to some extent to patterns observed in planetary atmosphere. For instance, the solid-body rotation corresponds roughly to the first order of the general circulation. This type of flows was also used in the past to model the phenomenon of super-rotation in planetary atmospheres \cite{Lim2007a}. The similarity is much more striking when considering the saddle points of the entropy functional (see \cite{Herbert2011d} for a detailed description), which although not formally stable, may be long-lived as the system may not generate spontaneously the destabilizing perturbations. In particular, we find states with vortices in the middle latitudes which are reminiscent of the atmosphere of the Earth. The lowest order saddle point is a quadrupole state, in agreement with simulations of 2D flows on a non-rotating sphere \cite{Cho1996,Marston2011}. In the rotating case, zonal structures prevail in numerical simulations \cite{Yoden1993,Cho1996}. When mixing is not strong enough, numerical simulations \cite{Brands1997} and laboratory experiments \cite{Marteau1995} indicate a discrepancy between the final organization of the flow and the statistical mechanics prediction. In this case, the system may remain in the vicinity of saddle points of the entropy functional for a long time.
Another example where the out-of-equilibrium steady-states are well described by the equilibrium theory \cite{Naso2010b} is that of the von Karman flow (the functions relating stream function, angular momentum and azimuthal velocity being selected by the forcing and dissipation \cite{Monchaux2006}). This indicates that the out-of-equilibrium attractors may concentrate near the equilibrium states and thus computations carried out in the framework of the equilibrium theory may remain relevant to describe the out-of-equilibrium steady-states when forcing and dissipation equilibrate each other.

%\bibliography{bibtexlib}

\begin{thebibliography}{54}
\expandafter\ifx\csname natexlab\endcsname\relax\def\natexlab#1{#1}\fi
\expandafter\ifx\csname bibnamefont\endcsname\relax
  \def\bibnamefont#1{#1}\fi
\expandafter\ifx\csname bibfnamefont\endcsname\relax
  \def\bibfnamefont#1{#1}\fi
\expandafter\ifx\csname citenamefont\endcsname\relax
  \def\citenamefont#1{#1}\fi
\expandafter\ifx\csname url\endcsname\relax
  \def\url#1{\texttt{#1}}\fi
\expandafter\ifx\csname urlprefix\endcsname\relax\def\urlprefix{URL }\fi
\providecommand{\bibinfo}[2]{#2}
\providecommand{\eprint}[2][]{\url{#2}}

\bibitem[{\citenamefont{McWilliams}(1984)}]{McWilliams1984}
\bibinfo{author}{\bibfnamefont{J.~C.} \bibnamefont{McWilliams}},
  \bibinfo{journal}{J. Fluid Mech.} \textbf{\bibinfo{volume}{146}},
  \bibinfo{pages}{21} (\bibinfo{year}{1984}).

\bibitem[{\citenamefont{Kraichnan}(1967)}]{Kraichnan1967}
\bibinfo{author}{\bibfnamefont{R.}~\bibnamefont{Kraichnan}},
  \bibinfo{journal}{Phys. Fluids} \textbf{\bibinfo{volume}{10}},
  \bibinfo{pages}{1417} (\bibinfo{year}{1967}).

\bibitem[{\citenamefont{Tabeling}(2002)}]{Tabeling2002}
\bibinfo{author}{\bibfnamefont{P.}~\bibnamefont{Tabeling}},
  \bibinfo{journal}{Phys. Rep.} \textbf{\bibinfo{volume}{362}},
  \bibinfo{pages}{1} (\bibinfo{year}{2002}).

\bibitem[{\citenamefont{Salmon et~al.}(1976)\citenamefont{Salmon, Holloway, and
  Hendershott}}]{Salmon1976}
\bibinfo{author}{\bibfnamefont{R.}~\bibnamefont{Salmon}},
  \bibinfo{author}{\bibfnamefont{G.}~\bibnamefont{Holloway}}, \bibnamefont{and}
  \bibinfo{author}{\bibfnamefont{M.}~\bibnamefont{Hendershott}},
  \bibinfo{journal}{J. Fluid Mech.} \textbf{\bibinfo{volume}{75}},
  \bibinfo{pages}{691} (\bibinfo{year}{1976}).

\bibitem[{\citenamefont{Frederiksen and Sawford}(1980)}]{Frederiksen1980}
\bibinfo{author}{\bibfnamefont{J.}~\bibnamefont{Frederiksen}} \bibnamefont{and}
  \bibinfo{author}{\bibfnamefont{B.}~\bibnamefont{Sawford}},
  \bibinfo{journal}{J. Atmos. Sci.} \textbf{\bibinfo{volume}{37}},
  \bibinfo{pages}{717} (\bibinfo{year}{1980}).

\bibitem[{\citenamefont{Miller}(1990)}]{Miller1990}
\bibinfo{author}{\bibfnamefont{J.}~\bibnamefont{Miller}},
  \bibinfo{journal}{Phys. Rev. Lett.} \textbf{\bibinfo{volume}{65}},
  \bibinfo{pages}{2137} (\bibinfo{year}{1990}).

\bibitem[{\citenamefont{Robert and Sommeria}(1991)}]{Robert1991a}
\bibinfo{author}{\bibfnamefont{R.}~\bibnamefont{Robert}} \bibnamefont{and}
  \bibinfo{author}{\bibfnamefont{J.}~\bibnamefont{Sommeria}},
  \bibinfo{journal}{J. Fluid Mech.} \textbf{\bibinfo{volume}{229}},
  \bibinfo{pages}{291} (\bibinfo{year}{1991}).

\bibitem[{\citenamefont{Robert}(1991)}]{Robert1991b}
\bibinfo{author}{\bibfnamefont{R.}~\bibnamefont{Robert}}, \bibinfo{journal}{J.
  Stat. Phys.} \textbf{\bibinfo{volume}{65}}, \bibinfo{pages}{531}
  (\bibinfo{year}{1991}).

\bibitem[{\citenamefont{Robert and Sommeria}(1992)}]{Robert1992}
\bibinfo{author}{\bibfnamefont{R.}~\bibnamefont{Robert}} \bibnamefont{and}
  \bibinfo{author}{\bibfnamefont{J.}~\bibnamefont{Sommeria}},
  \bibinfo{journal}{Phys. Rev. Lett.} \textbf{\bibinfo{volume}{69}},
  \bibinfo{pages}{2776} (\bibinfo{year}{1992}).

\bibitem[{\citenamefont{Kazantsev et~al.}(1998)\citenamefont{Kazantsev,
  Sommeria, and Verron}}]{Kazantsev1998}
\bibinfo{author}{\bibfnamefont{E.}~\bibnamefont{Kazantsev}},
  \bibinfo{author}{\bibfnamefont{J.}~\bibnamefont{Sommeria}}, \bibnamefont{and}
  \bibinfo{author}{\bibfnamefont{J.}~\bibnamefont{Verron}},
  \bibinfo{journal}{J. Phys. Oceanogr.} \textbf{\bibinfo{volume}{28}},
  \bibinfo{pages}{1017} (\bibinfo{year}{1998}).

\bibitem[{\citenamefont{Polyakov}(2001)}]{Polyakov2001}
\bibinfo{author}{\bibfnamefont{I.}~\bibnamefont{Polyakov}},
  \bibinfo{journal}{J. Phys. Oceanogr.} \textbf{\bibinfo{volume}{31}},
  \bibinfo{pages}{2255} (\bibinfo{year}{2001}).

\bibitem[{\citenamefont{Chavanis and Sommeria}(1997)}]{Chavanis1997}
\bibinfo{author}{\bibfnamefont{P.-H.} \bibnamefont{Chavanis}} \bibnamefont{and}
  \bibinfo{author}{\bibfnamefont{J.}~\bibnamefont{Sommeria}},
  \bibinfo{journal}{Phys. Rev. Lett.} \textbf{\bibinfo{volume}{78}},
  \bibinfo{pages}{3302} (\bibinfo{year}{1997}).

\bibitem[{\citenamefont{Chavanis and Sommeria}(2002)}]{Chavanis2002a}
\bibinfo{author}{\bibfnamefont{P.-H.} \bibnamefont{Chavanis}} \bibnamefont{and}
  \bibinfo{author}{\bibfnamefont{J.}~\bibnamefont{Sommeria}},
  \bibinfo{journal}{Phys. Rev. E} \textbf{\bibinfo{volume}{65}},
  \bibinfo{pages}{26302} (\bibinfo{year}{2002}).

\bibitem[{\citenamefont{Chavanis}(2009)}]{Chavanis2009}
\bibinfo{author}{\bibfnamefont{P.-H.} \bibnamefont{Chavanis}},
  \bibinfo{journal}{Eur. Phys. J. B} \textbf{\bibinfo{volume}{70}},
  \bibinfo{pages}{73} (\bibinfo{year}{2009}).

\bibitem[{\citenamefont{Michel and Robert}(1994)}]{Michel1994a}
\bibinfo{author}{\bibfnamefont{J.}~\bibnamefont{Michel}} \bibnamefont{and}
  \bibinfo{author}{\bibfnamefont{R.}~\bibnamefont{Robert}},
  \bibinfo{journal}{J. Stat. Phys.} \textbf{\bibinfo{volume}{77}},
  \bibinfo{pages}{645} (\bibinfo{year}{1994}).

\bibitem[{\citenamefont{Turkington et~al.}(2001)\citenamefont{Turkington,
  Majda, Haven, and DiBattista}}]{Turkington2001}
\bibinfo{author}{\bibfnamefont{B.}~\bibnamefont{Turkington}},
  \bibinfo{author}{\bibfnamefont{A.}~\bibnamefont{Majda}},
  \bibinfo{author}{\bibfnamefont{K.}~\bibnamefont{Haven}}, \bibnamefont{and}
  \bibinfo{author}{\bibfnamefont{M.}~\bibnamefont{DiBattista}},
  \bibinfo{journal}{Proc. Natl. Acad. Sci. U.S.A.}
  \textbf{\bibinfo{volume}{98}}, \bibinfo{pages}{12346} (\bibinfo{year}{2001}).

\bibitem[{\citenamefont{Bouchet and Sommeria}(2002)}]{Bouchet2002}
\bibinfo{author}{\bibfnamefont{F.}~\bibnamefont{Bouchet}} \bibnamefont{and}
  \bibinfo{author}{\bibfnamefont{J.}~\bibnamefont{Sommeria}},
  \bibinfo{journal}{J. Fluid Mech.} \textbf{\bibinfo{volume}{464}},
  \bibinfo{pages}{165} (\bibinfo{year}{2002}).

\bibitem[{\citenamefont{Weichman}(2006)}]{Weichman2006}
\bibinfo{author}{\bibfnamefont{P.}~\bibnamefont{Weichman}},
  \bibinfo{journal}{Phys. Rev. E} \textbf{\bibinfo{volume}{73}},
  \bibinfo{pages}{36313} (\bibinfo{year}{2006}).

\bibitem[{\citenamefont{Venaille and Bouchet}()}]{VenailleArxiv}
\bibinfo{author}{\bibfnamefont{A.}~\bibnamefont{Venaille}} \bibnamefont{and}
  \bibinfo{author}{\bibfnamefont{F.}~\bibnamefont{Bouchet}},
  \bibinfo{journal}{J. Phys. Oceanogr.} \textbf{\bibinfo{volume}{41}},
  \bibinfo{pages}{1860} (\bibinfo{year}{2011}).

\bibitem[{\citenamefont{Venaille and Bouchet}(2009)}]{Venaille2009}
\bibinfo{author}{\bibfnamefont{A.}~\bibnamefont{Venaille}} \bibnamefont{and}
  \bibinfo{author}{\bibfnamefont{F.}~\bibnamefont{Bouchet}},
  \bibinfo{journal}{Phys. Rev. Lett.} \textbf{\bibinfo{volume}{102}},
  \bibinfo{pages}{104501} (\bibinfo{year}{2009}).

\bibitem[{\citenamefont{Venaille and Bouchet}(2011)}]{Venaille2011}
\bibinfo{author}{\bibfnamefont{A.}~\bibnamefont{Venaille}} \bibnamefont{and}
  \bibinfo{author}{\bibfnamefont{F.}~\bibnamefont{Bouchet}},
  \bibinfo{journal}{J. Stat. Phys.} \textbf{\bibinfo{volume}{143}},
  \bibinfo{pages}{346} (\bibinfo{year}{2011}).

\bibitem[{\citenamefont{Naso et~al.}(2011)\citenamefont{Naso, Chavanis, and
  Dubrulle}}]{Naso2011}
\bibinfo{author}{\bibfnamefont{A.}~\bibnamefont{Naso}},
  \bibinfo{author}{\bibfnamefont{P.-H.} \bibnamefont{Chavanis}},
  \bibnamefont{and} \bibinfo{author}{\bibfnamefont{B.}~\bibnamefont{Dubrulle}},
  \bibinfo{journal}{Eur. Phys. J. B} \textbf{\bibinfo{volume}{80}},
  \bibinfo{pages}{493} (\bibinfo{year}{2011}).

\bibitem[{\citenamefont{Dauxois et~al.}(2002)\citenamefont{Dauxois, Ruffo,
  Arimondo, and Wilkens}}]{DauxoisLRIbook}
\bibinfo{editor}{\bibfnamefont{T.}~\bibnamefont{Dauxois}},
  \bibinfo{editor}{\bibfnamefont{S.}~\bibnamefont{Ruffo}},
  \bibinfo{editor}{\bibfnamefont{E.}~\bibnamefont{Arimondo}}, \bibnamefont{and}
  \bibinfo{editor}{\bibfnamefont{M.}~\bibnamefont{Wilkens}}, eds.,
  \emph{\bibinfo{title}{Dynamics and Thermodynamics of Systems with Long Range
  Interactions}}, vol. \bibinfo{volume}{602} of \emph{\bibinfo{series}{Lecture
  Notes in Physics}} (\bibinfo{publisher}{Springer, New-York},
  \bibinfo{year}{2002}).

\bibitem[{\citenamefont{Campa et~al.}(2009)\citenamefont{Campa, Dauxois, and
  Ruffo}}]{Campa2009}
\bibinfo{author}{\bibfnamefont{A.}~\bibnamefont{Campa}},
  \bibinfo{author}{\bibfnamefont{T.}~\bibnamefont{Dauxois}}, \bibnamefont{and}
  \bibinfo{author}{\bibfnamefont{S.}~\bibnamefont{Ruffo}},
  \bibinfo{journal}{Phys. Rep.} \textbf{\bibinfo{volume}{480}},
  \bibinfo{pages}{57} (\bibinfo{year}{2009}).

\bibitem[{\citenamefont{Lynden-Bell and Wood}(1968)}]{LyndenBell1968}
\bibinfo{author}{\bibfnamefont{D.}~\bibnamefont{Lynden-Bell}} \bibnamefont{and}
  \bibinfo{author}{\bibfnamefont{R.}~\bibnamefont{Wood}},
  \bibinfo{journal}{Mon. Not. R. Astron. Soc.} \textbf{\bibinfo{volume}{138}},
  \bibinfo{pages}{495} (\bibinfo{year}{1968}).

\bibitem[{\citenamefont{Thirring}(1970)}]{Thirring1970}
\bibinfo{author}{\bibfnamefont{W.}~\bibnamefont{Thirring}},
  \bibinfo{journal}{Z. Physik.} \textbf{\bibinfo{volume}{235}},
  \bibinfo{pages}{339} (\bibinfo{year}{1970}).

\bibitem[{\citenamefont{Ellis et~al.}(2000)\citenamefont{Ellis, Haven, and
  Turkington}}]{Ellis2000}
\bibinfo{author}{\bibfnamefont{R.}~\bibnamefont{Ellis}},
  \bibinfo{author}{\bibfnamefont{K.}~\bibnamefont{Haven}}, \bibnamefont{and}
  \bibinfo{author}{\bibfnamefont{B.}~\bibnamefont{Turkington}},
  \bibinfo{journal}{J. Stat. Phys.} \textbf{\bibinfo{volume}{101}},
  \bibinfo{pages}{999} (\bibinfo{year}{2000}).

\bibitem[{\citenamefont{Bouchet and Barre}(2005)}]{Bouchet2005}
\bibinfo{author}{\bibfnamefont{F.}~\bibnamefont{Bouchet}} \bibnamefont{and}
  \bibinfo{author}{\bibfnamefont{J.}~\bibnamefont{Barre}}, \bibinfo{journal}{J.
  Stat. Phys.} \textbf{\bibinfo{volume}{118}}, \bibinfo{pages}{1073}
  (\bibinfo{year}{2005}).

\bibitem[{\citenamefont{Chavanis}(2006)}]{Chavanis2006h}
\bibinfo{author}{\bibfnamefont{P.-H.} \bibnamefont{Chavanis}},
  \bibinfo{journal}{Int. J. Mod. Phys. B} \textbf{\bibinfo{volume}{20}},
  \bibinfo{pages}{3113} (\bibinfo{year}{2006}).

\bibitem[{\citenamefont{Ding and Lim}(2007)}]{Ding2007}
\bibinfo{author}{\bibfnamefont{X.}~\bibnamefont{Ding}} \bibnamefont{and}
  \bibinfo{author}{\bibfnamefont{C.~C.} \bibnamefont{Lim}},
  \bibinfo{journal}{Physica A} \textbf{\bibinfo{volume}{374}},
  \bibinfo{pages}{152} (\bibinfo{year}{2007}).

\bibitem[{\citenamefont{Lim}(2007)}]{Lim2007a}
\bibinfo{author}{\bibfnamefont{C.~C.} \bibnamefont{Lim}}, \bibinfo{journal}{J.
  Math. Phys.} \textbf{\bibinfo{volume}{48}}, \bibinfo{pages}{065603}
  (\bibinfo{year}{2007}).

\bibitem[{\citenamefont{Majda and Wang}(2006)}]{MajdaWangBook}
\bibinfo{author}{\bibfnamefont{A.}~\bibnamefont{Majda}} \bibnamefont{and}
  \bibinfo{author}{\bibfnamefont{X.}~\bibnamefont{Wang}},
  \emph{\bibinfo{title}{{Nonlinear Dynamics and Statistical Theories for Basic
  Geophysical Flows}}} (\bibinfo{publisher}{Cambridge University Press},
  \bibinfo{address}{Cambridge}, \bibinfo{year}{2006}).

\bibitem[{\citenamefont{Herbert et~al.}(2012)\citenamefont{Herbert, Dubrulle,
  Chavanis, and Paillard}}]{Herbert2011d}
\bibinfo{author}{\bibfnamefont{C.}~\bibnamefont{Herbert}},
  \bibinfo{author}{\bibfnamefont{B.}~\bibnamefont{Dubrulle}},
  \bibinfo{author}{\bibfnamefont{P.~H.} \bibnamefont{Chavanis}},
  \bibnamefont{and} \bibinfo{author}{\bibfnamefont{D.}~\bibnamefont{Paillard}},
  \bibinfo{journal}{J. Stat. Mech., submitted}  (\bibinfo{year}{2012}).

\bibitem[{\citenamefont{Bouchet}(2008)}]{Bouchet2008}
\bibinfo{author}{\bibfnamefont{F.}~\bibnamefont{Bouchet}},
  \bibinfo{journal}{Physica D} \textbf{\bibinfo{volume}{237}},
  \bibinfo{pages}{1976} (\bibinfo{year}{2008}).

\bibitem[{\citenamefont{Chavanis and Sommeria}(1996)}]{Chavanis1996}
\bibinfo{author}{\bibfnamefont{P.-H.} \bibnamefont{Chavanis}} \bibnamefont{and}
  \bibinfo{author}{\bibfnamefont{J.}~\bibnamefont{Sommeria}},
  \bibinfo{journal}{J. Fluid Mech.} \textbf{\bibinfo{volume}{314}},
  \bibinfo{pages}{267} (\bibinfo{year}{1996}).

\bibitem[{\citenamefont{Naso et~al.}(2010{\natexlab{a}})\citenamefont{Naso,
  Chavanis, and Dubrulle}}]{Naso2010a}
\bibinfo{author}{\bibfnamefont{A.}~\bibnamefont{Naso}},
  \bibinfo{author}{\bibfnamefont{P.-H.} \bibnamefont{Chavanis}},
  \bibnamefont{and} \bibinfo{author}{\bibfnamefont{B.}~\bibnamefont{Dubrulle}},
  \bibinfo{journal}{Eur. Phys. J. B} \textbf{\bibinfo{volume}{77}},
  \bibinfo{pages}{187} (\bibinfo{year}{2010}{\natexlab{a}}).

\bibitem[{\citenamefont{Ellis et~al.}(2002)\citenamefont{Ellis, Haven, and
  Turkington}}]{Ellis2002}
\bibinfo{author}{\bibfnamefont{R.}~\bibnamefont{Ellis}},
  \bibinfo{author}{\bibfnamefont{K.}~\bibnamefont{Haven}}, \bibnamefont{and}
  \bibinfo{author}{\bibfnamefont{B.}~\bibnamefont{Turkington}},
  \bibinfo{journal}{Nonlinearity} \textbf{\bibinfo{volume}{15}},
  \bibinfo{pages}{239} (\bibinfo{year}{2002}).

\bibitem[{\citenamefont{Chavanis}(2008)}]{Chavanis2008b}
\bibinfo{author}{\bibfnamefont{P.-H.} \bibnamefont{Chavanis}},
  \bibinfo{journal}{Physica D} \textbf{\bibinfo{volume}{237}},
  \bibinfo{pages}{1998} (\bibinfo{year}{2008}).

\bibitem[{\citenamefont{Chavanis et~al.}(2010)\citenamefont{Chavanis, Naso, and
  Dubrulle}}]{Chavanis2010c}
\bibinfo{author}{\bibfnamefont{P.-H.} \bibnamefont{Chavanis}},
  \bibinfo{author}{\bibfnamefont{A.}~\bibnamefont{Naso}}, \bibnamefont{and}
  \bibinfo{author}{\bibfnamefont{B.}~\bibnamefont{Dubrulle}},
  \bibinfo{journal}{Eur. Phys. J. B} \textbf{\bibinfo{volume}{77}},
  \bibinfo{pages}{167} (\bibinfo{year}{2010}).

\bibitem[{\citenamefont{Touchette et~al.}(2004)\citenamefont{Touchette, Ellis,
  and Turkington}}]{Touchette2004}
\bibinfo{author}{\bibfnamefont{H.}~\bibnamefont{Touchette}},
  \bibinfo{author}{\bibfnamefont{R.~S.} \bibnamefont{Ellis}}, \bibnamefont{and}
  \bibinfo{author}{\bibfnamefont{B.}~\bibnamefont{Turkington}},
  \bibinfo{journal}{Physica A} \textbf{\bibinfo{volume}{340}},
  \bibinfo{pages}{138} (\bibinfo{year}{2004}).

\bibitem[{\citenamefont{Kraichnan}(1975)}]{Kraichnan1975}
\bibinfo{author}{\bibfnamefont{R.}~\bibnamefont{Kraichnan}},
  \bibinfo{journal}{J. Fluid Mech.} \textbf{\bibinfo{volume}{67}},
  \bibinfo{pages}{155} (\bibinfo{year}{1975}).

\bibitem[{\citenamefont{Carnevale and Frederiksen}(1987)}]{Carnevale1987}
\bibinfo{author}{\bibfnamefont{G.}~\bibnamefont{Carnevale}} \bibnamefont{and}
  \bibinfo{author}{\bibfnamefont{J.}~\bibnamefont{Frederiksen}},
  \bibinfo{journal}{J. Fluid Mech.} \textbf{\bibinfo{volume}{175}},
  \bibinfo{pages}{157} (\bibinfo{year}{1987}).

\bibitem[{\citenamefont{Zinn-Justin}(2002)}]{ZinnJustinBook}
\bibinfo{author}{\bibfnamefont{J.}~\bibnamefont{Zinn-Justin}},
  \emph{\bibinfo{title}{Quantum Field Theory and Critical Phenomena}}
  (\bibinfo{publisher}{Oxford University Press}, \bibinfo{year}{2002}).

\bibitem[{\citenamefont{Bouchet and Corvellec}(2010)}]{Bouchet2010}
\bibinfo{author}{\bibfnamefont{F.}~\bibnamefont{Bouchet}} \bibnamefont{and}
  \bibinfo{author}{\bibfnamefont{M.}~\bibnamefont{Corvellec}},
  \bibinfo{journal}{J. Stat. Mech.} \textbf{\bibinfo{volume}{2010}},
  \bibinfo{pages}{P08021} (\bibinfo{year}{2010}).

\bibitem[{\citenamefont{Corvellec}(2012)}]{CorvellecThesis}
\bibinfo{author}{\bibfnamefont{M.}~\bibnamefont{Corvellec}}, Ph.D. thesis,
  \bibinfo{school}{ENS Lyon} (\bibinfo{year}{2012}).

\bibitem[{\citenamefont{Bouchet and Venaille}(2012)}]{Bouchet2012}
\bibinfo{author}{\bibfnamefont{F.}~\bibnamefont{Bouchet}} \bibnamefont{and}
  \bibinfo{author}{\bibfnamefont{A.}~\bibnamefont{Venaille}},
  \bibinfo{journal}{Phys. Rep.} \textbf{\bibinfo{volume}{in press}}
  (\bibinfo{year}{2012}).

\bibitem[{\citenamefont{Pedlosky}(1987)}]{PedloskyGFD}
\bibinfo{author}{\bibfnamefont{J.}~\bibnamefont{Pedlosky}},
  \emph{\bibinfo{title}{Geophysical Fluid Dynamics}}
  (\bibinfo{publisher}{Springer}, \bibinfo{year}{1987}).

\bibitem[{\citenamefont{Cho and Polvani}(1996)}]{Cho1996}
\bibinfo{author}{\bibfnamefont{J.}~\bibnamefont{Cho}} \bibnamefont{and}
  \bibinfo{author}{\bibfnamefont{L.}~\bibnamefont{Polvani}},
  \bibinfo{journal}{Phys. Fluids} \textbf{\bibinfo{volume}{8}},
  \bibinfo{pages}{1531} (\bibinfo{year}{1996}).

\bibitem[{\citenamefont{Marston}(2011)}]{Marston2011}
\bibinfo{author}{\bibfnamefont{B.}~\bibnamefont{Marston}},
  \bibinfo{journal}{Physics} \textbf{\bibinfo{volume}{4}}, \bibinfo{pages}{20}
  (\bibinfo{year}{2011}).

\bibitem[{\citenamefont{Yoden and Yamada}(1993)}]{Yoden1993}
\bibinfo{author}{\bibfnamefont{S.}~\bibnamefont{Yoden}} \bibnamefont{and}
  \bibinfo{author}{\bibfnamefont{M.}~\bibnamefont{Yamada}},
  \bibinfo{journal}{J. Atmos. Sci.} \textbf{\bibinfo{volume}{50}},
  \bibinfo{pages}{631} (\bibinfo{year}{1993}).

\bibitem[{\citenamefont{Brands et~al.}(1997)\citenamefont{Brands, Stulemeyer,
  Pasmanter, and Schep}}]{Brands1997}
\bibinfo{author}{\bibfnamefont{H.}~\bibnamefont{Brands}},
  \bibinfo{author}{\bibfnamefont{J.}~\bibnamefont{Stulemeyer}},
  \bibinfo{author}{\bibfnamefont{R.~A.} \bibnamefont{Pasmanter}},
  \bibnamefont{and} \bibinfo{author}{\bibfnamefont{T.}~\bibnamefont{Schep}},
  \bibinfo{journal}{Phys. Fluids} \textbf{\bibinfo{volume}{9}},
  \bibinfo{pages}{2815} (\bibinfo{year}{1997}).

\bibitem[{\citenamefont{Marteau et~al.}(1995)\citenamefont{Marteau, Cardoso,
  and Tabeling}}]{Marteau1995}
\bibinfo{author}{\bibfnamefont{D.}~\bibnamefont{Marteau}},
  \bibinfo{author}{\bibfnamefont{O.}~\bibnamefont{Cardoso}}, \bibnamefont{and}
  \bibinfo{author}{\bibfnamefont{P.}~\bibnamefont{Tabeling}},
  \bibinfo{journal}{Phys. Rev. E} \textbf{\bibinfo{volume}{51}},
  \bibinfo{pages}{5124} (\bibinfo{year}{1995}).

\bibitem[{\citenamefont{Naso et~al.}(2010{\natexlab{b}})\citenamefont{Naso,
  Monchaux, Chavanis, and Dubrulle}}]{Naso2010b}
\bibinfo{author}{\bibfnamefont{A.}~\bibnamefont{Naso}},
  \bibinfo{author}{\bibfnamefont{R.}~\bibnamefont{Monchaux}},
  \bibinfo{author}{\bibfnamefont{P.-H.} \bibnamefont{Chavanis}},
  \bibnamefont{and} \bibinfo{author}{\bibfnamefont{B.}~\bibnamefont{Dubrulle}},
  \bibinfo{journal}{Phys. Rev. E} \textbf{\bibinfo{volume}{81}},
  \bibinfo{pages}{066318} (\bibinfo{year}{2010}{\natexlab{b}}).

\bibitem[{\citenamefont{Monchaux et~al.}(2006)\citenamefont{Monchaux, Ravelet,
  Dubrulle, Chiffaudel, and Daviaud}}]{Monchaux2006}
\bibinfo{author}{\bibfnamefont{R.}~\bibnamefont{Monchaux}},
  \bibinfo{author}{\bibfnamefont{F.}~\bibnamefont{Ravelet}},
  \bibinfo{author}{\bibfnamefont{B.}~\bibnamefont{Dubrulle}},
  \bibinfo{author}{\bibfnamefont{A.}~\bibnamefont{Chiffaudel}},
  \bibnamefont{and} \bibinfo{author}{\bibfnamefont{F.}~\bibnamefont{Daviaud}},
  \bibinfo{journal}{Phys. Rev. Lett.} \textbf{\bibinfo{volume}{96}},
  \bibinfo{pages}{124502} (\bibinfo{year}{2006}).

\end{thebibliography}

\end{document}